\documentclass{aa}
\newcounter{appendix2}
\renewcommand{\thefigure}{\arabic{figure}\Alph{appendix2}}
\usepackage{graphicx}
\usepackage{amsmath}
\begin{document}
\title{Initial magnetic field configurations for 3-dimensional simulations
of astrophysical jets}
\titlerunning{Magnetic fields in 3-D simulations of astrophysical jets}
\author{Michael A. S. G. J{\o}rgensen\inst{1}, Rachid Ouyed\inst{2}
\and Morten Christensen\inst{1}}
\institute{$^{1}$University of Copenhagen, Astronomical Observatory, Juliane
Maries Vej 30, DK-2100 Copenhagen {\O}, Denmark\\
$^{2}$Nordic Institute for Theoretical Physics, Blegdamsvej 17, DK-2100
Copenhagen {\O}, Denmark}
\date{Received/Accepted}
\authorrunning{J{\o}rgensen, Ouyed \& Christensen}

\offprints{M. J{\o}rgensen,\\ \email{svart@astro.ku.dk}} 

\date{Received ?? 2001/Accepted ?? 2001}

\abstract{
We  solve, and provide analytical expressions, for current-free magnetic
configurations in the context of initial setups of 3-dimensional
simulations of astrophysical jets involving
an accretion disk corona in hydrostatic balance around
a central object. These configurations which thread through
the accretion disk and its corona preserve the initial hydrostatic state. 
This work sets stage for future  3-dimensional jet simulations
(including disk rotation and mass-load) where
 launching, acceleration  and collimation mechanisms
can be investigated.
}

\maketitle

\section{Introduction}

The most successful model of astrophysical jets is that
of magneto-hydrodynamic (MHD) driven winds as presented by Blandford \& Payne (1982).
In this model, the magnetic field is predicted to
launch, accelerate and  efficiently collimate the jet.
 The advent of MHD codes
over the last decade allowed us to study the details of this model
through the use of time-dependent MHD simulations
and confirm the importance of the magnetic field. 
One point that has been emphasized in many of these
simulations is the 
importance of
properly setting up the initial state. A numerically
stable initial setup makes the simulation tractable 
and allows contact with theory. 
Here, we are particularly interested
in simulations as presented  
in Ouyed \& Pudritz (1997a, OPI) and Ouyed \& Pudritz  (1997b, OPII)
focusing on the technical aspect of the initial state.

The initial conditions as defined in OPI and OPII
correspond to  a central object (a proto-star) 
surrounded by a Keplerian disk and an overlying corona in
hydrostatic equilibrium. The disk has fixed properties and provides the  
boundary conditions for the outflow.
The time evolution of the initial state is
partly depicted by the momentum equation (using the standard nomenclature),  
\begin{equation}
\rho \left({\partial{\bf v}\over \partial t}+({\bf v\cdot \nabla}){\bf v}\right)
+\nabla p +\rho {\bf \nabla}\Phi -
{\bf J}\times {\bf B}=0,
\end{equation}
and would, in general, be perturbed 
by the Lorentz force of a  magnetic field that threads through
the accretion disk and its corona.  Therefore, 
all previous 2-dimensional (2-D) simulations (OPI\&II) employ  
initialy current-free, ${\bf J}=0$, magnetic fields
where the condition for  stability becomes
\begin{equation}
0=-{\nabla p\over\rho}-{\bf \nabla}\Phi,
\end{equation}
at all time.
This initial setup
was successfully extended to 3-dimensional (3-D) simulations in Ouyed, Pudritz,
\& Clarke (2001, OPC). However, the work presented in OPC was restricted to the
simple case of a uniform magnetic configuration where the
accretion  disks is threaded by vertical field lines. 
In this paper, we solve for non-uniform current-free magnetic fields in 3-D 
which would complete our 3-D initial hydrostatic corona as defined in (eq~(2)). 
The paper is organized as follows:   
In \S 2, we describe the force-free
equations defining our problem. In  \S 3 we solve
for the current-free case and test the stability of these
solutions by evolving the initial setup in time. We introduce and 
describe the {\it JETSET} tool developed for this purpose. We conclude  in 
\S 4.

\section{Force-free  configurations}

Force-free configurations
 are characterized by 
\begin{equation}
(\vec{J}\times\vec{B} = 0, 
\vec{\nabla}\cdot\vec{B} = 0),
\end{equation}
corresponding to the standard two cases,
$\vec{J}\|\vec{B}$, and $\vec{J} = 0$.
We start with the case where currents are parallel to the magnetic
field lines. Or,
\begin{equation}
 \vec{B} = \vec{\nabla}\times\left (\frac{\vec{B}}{\alpha}\right ),
\label{3Dff} 
\end{equation}
(here, $\alpha$ is constant), i.e. the vector potential is given by
$\frac{1}{\alpha}\vec{B}$ (from $\vec{B}=\vec{\nabla}\times\vec{A}$). Thus we
first solve for the magnetic field and then calculate the vector potential;
recall that in deriving {\bf B} using the vector potential, 
$\vec{\nabla}\cdot\vec{B}=0$ is  numerically guaranteed to within machine
round-off errors. Here and in the rest of the paper, we adopt cylindrical
coordinates (r,$\phi$,z) and we  assume that the fields are separable.

Using relation
 \eqref{3Dff}  and  the fact that the magnetic field is
divergence free we arrive at:
\begin{equation}
\alpha^2\vec{B} = 
-\vec{\nabla}^2\vec{B}.
 \end{equation}
This is a vector relation and therefore also holds 
for every component. Let us write a component of the magnetic field, $\vec{B}$,
as $B$ and making use of our usual assumption of separable variables we find: 
\begin{eqnarray}
 B&=&R(r) \Phi(\phi)  Z(z), \nonumber\\
R&=&D(\rho,\lambda)  J_\rho\left (\sqrt{\alpha^2+\lambda ^2}  r\right ), \nonumber\\
\Phi &=& E(\rho,\lambda) \sin(\rho\phi)+C(\rho,\lambda) \cos(\rho\phi), \nonumber\\
Z &=& A(\rho,\lambda)   e^{-\lambda |z|}.
\end{eqnarray}
where $\lambda$ and $\rho$
are constants of separation, and the unknown functions $A, C, D, E$ are to be
determined from the boundary conditions.
The complete solution for $B$ is the integral over
$\lambda$ ranging from 0 to infinity, and a sum over every $\rho$
(Arfken \& Weber 1995):
\begin{eqnarray}
B(r,\phi,z) &=& \sum_{\rho=0}^\infty \left[ \int_0^\infty
 D(\rho,\lambda) J_\rho 
\left ( 
\sqrt{\alpha^2+\lambda ^2} r
\right )
  A(\rho,\lambda)\right.\nonumber \\
&\times&
\left.
e^{-\lambda |z|} 
  \left ( 
E(\rho,\lambda) \sin(\rho\phi)+C(\rho,\lambda) \cos(\rho\phi) 
  \right )
\right ]\nonumber.
\end{eqnarray}
For every component one finds 
two constants given through Fourier expansion. That is, extra
boundary conditions are required. In our case for example, all components of
the magnetic field in the accretion disk must be specified. However, the integral
equations remain difficult to treat  since the Bessel function is not linearly
dependent on $\lambda$, and analytical solutions are not straightforward. We
thus turn to the current-free case ($\alpha=0$) where the equations can
be simplified.

\section{Current-free configurations}

Here, $\vec{J} = \vec{0}$ is a necessary and sufficient condition to
guaranty the existence of a scalar field ($\varphi$) with the property,
$\vec{\nabla} \varphi = \vec{B}$. As can be seen in appendix \ref{cfapp} this
greatly simplifies our task of finding the initial configurations.
We investigate two kind of current-free magnetic fields, with and without
toroidal component ($B_{\phi}$).  

\subsection{$B_{\phi}\ne 0$}

Here  we do not demand that the toroidal field is zero in the corona.  We use 
 the following boundary condition:
\begin{equation}\label{bond2}
B_{\phi,0} = (b\ r^{\mu-1})\times \sin(k\phi),
\end{equation}
where $B_{\phi,0}$ is the toroidal
 magnetic field in the disk, and $b$ 
a normalization factor. The
general solution for the scalar field is
given  in appendix \ref{cfapp} where we
also explain the choice of such a $\phi$ dependence.
We find,
\begin{eqnarray}
\varphi(r,\phi,z) &=& q(\mu,k) b
 \cos(k\phi)\frac{r^k}{\sqrt{r^2+|z|^2}^{k-\mu+2}} \hfill \nonumber \\
&\hfill & \times\  _2F_1 \left (\frac{k-\mu +1}{2},
 \frac{\mu+k}{2},k+1,\frac{r^2}{r^2+|z|^2}\right ),
\label{cfphi}\nonumber \\
 q(\mu,k) &=& 2^{\mu-k+1}\frac{\Gamma\left (\frac{1}{2}(k+\mu+2)\right )
\Gamma\left (k-\mu+2\right )}
{\Gamma\left (\frac{1}{2}(k-\mu)\right )\Gamma\left (k+1\right )},\nonumber\\
&&\quad \hbox{for}\quad 1-k < \mu < \frac{1}{2}\ , 
\end{eqnarray}
where $_2F_1$ is a hyper geometric 
function and $\Gamma$ the gamma function.
The $k=2$ case, for example, is a simple solution. The corresponding
configurations once implemented in the simulations remain stable in time. 
In the $B_{\phi}\ne 0$ case, however, the corresponding jet 
simulations (including mass-load and disk rotation) are prone to pinch forces  
within few disk rotations. That is, the resulting dynamics is more
complex, and the jet more
difficult to track/investigate numerically. 
The $B_{\phi}=0$ configurations,  also adopted in OPI\&II,
turned out to be useful in many ways; it allows for instance to demonstrate
that the jet collimation can result from the self-generated toroidal field.
These we consider in details next.

\subsection{$B_{\phi}=0$}

In the current-free initial setup the differential
 equation for the scalar field of the magnetic field ($\varphi$) reduces to
the Laplace equation (since $\vec{\nabla} \cdot \vec{B} = 0$):
\begin{equation}\label{laplace}
\vec{\nabla}^2 \varphi(r,\phi,z) = 0.
\end{equation}
The corresponding scalar potential is found to be (Appendix \ref{cfapp}),
\begin{eqnarray} \label{cfnophi}
\varphi(r,\phi,z) &=& 2^\mu b  z^{\mu}
\frac{\Gamma(1+\frac{\mu}{2})\Gamma(-\mu)}{\Gamma(1-\frac{\mu}{2})}
{}_2F_1(\frac{1-\mu}{2},-\frac{\mu}{2},1,-\frac{r^2}{z^2}),\nonumber\\
 & \hfill & -2 < \mu < \frac{1}{2}\ .
\end{eqnarray}
Solution above corresponds to the following 
boundary condition:
\begin{equation}\label{bond}
B_{r,0} = b\ r^{\mu-1},
\end{equation}
where $B_{r,0}$ is the radial component of
 the magnetic field in the accretion disk.
 Recall that we do not simulate the accretion disk
(a fixed boundary in all of our simulations), but 
only the jet and the overlying corona ($z > 0$). 

We examined configurations for  2
 different values of $\mu$ (the radial dependence of the magnetic field); 
an open ($\mu = -1$) and a closed
($\mu = -2$) configuration. The
closed configuration  has the same radial
dependence as a dipole field, which is expected (in first approximation)
around stars however ours is strictly a solution of the
disk boundary condition.  The open
configuration is the 3-D  analogue/extension of the open configuration
used in OPI.
We implemented these 2 configurations
in our initial setup and let them evolve in time after applying
a small perturbation to the  density.
No mass injection and no disk rotation were set,
and  $v_{z}=v_{r}=v_{\phi}=0$ in the corona ($z > 0.0$).
For the simulations we used the time-explicit 
Eulerian MHD-code ZEUS3D (Stone \& Norman 1992).

\begin{figure}[t!]
\begin{center}
 \resizebox{6cm}{!}{\includegraphics{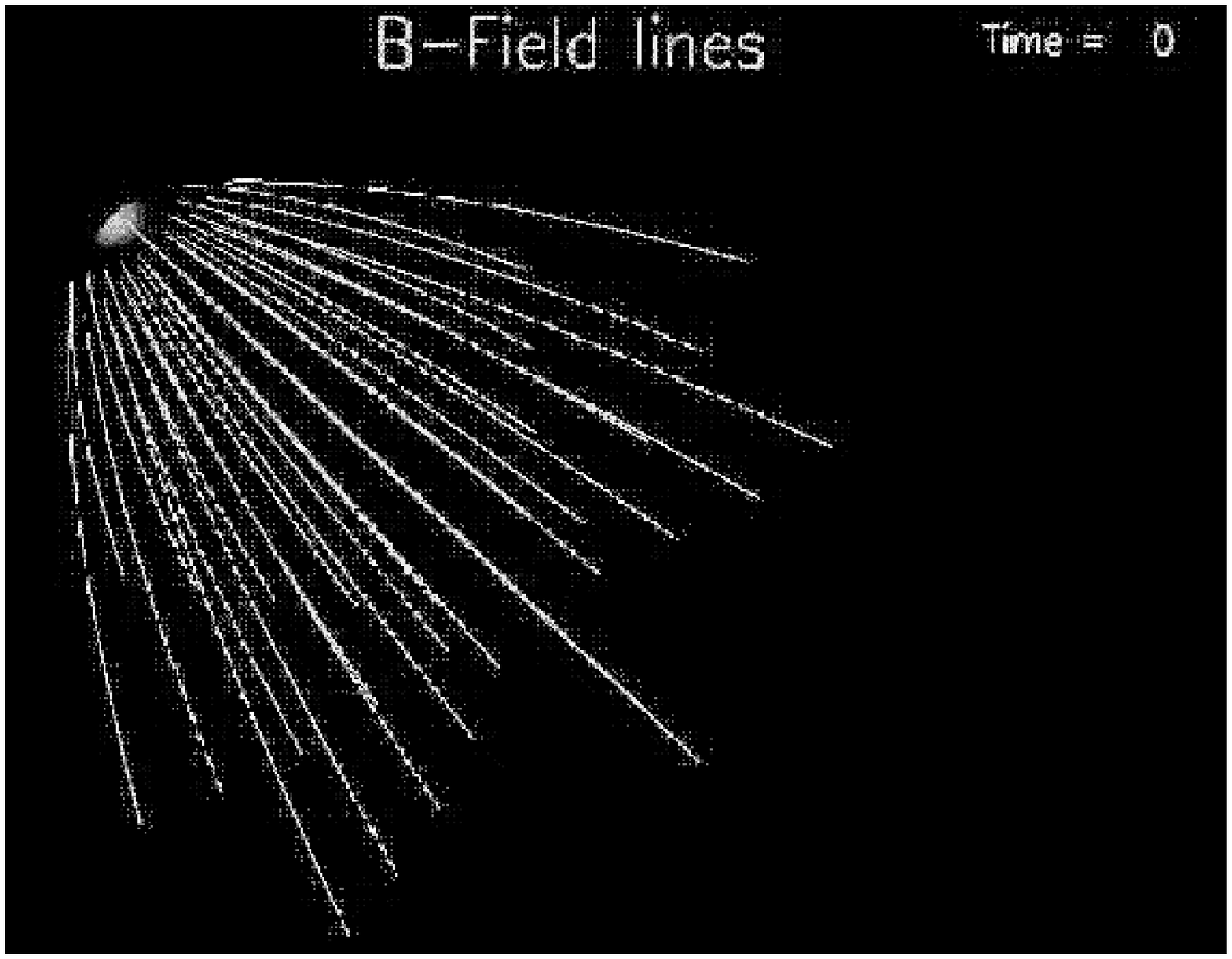}}
 \resizebox{6cm}{!}{\includegraphics{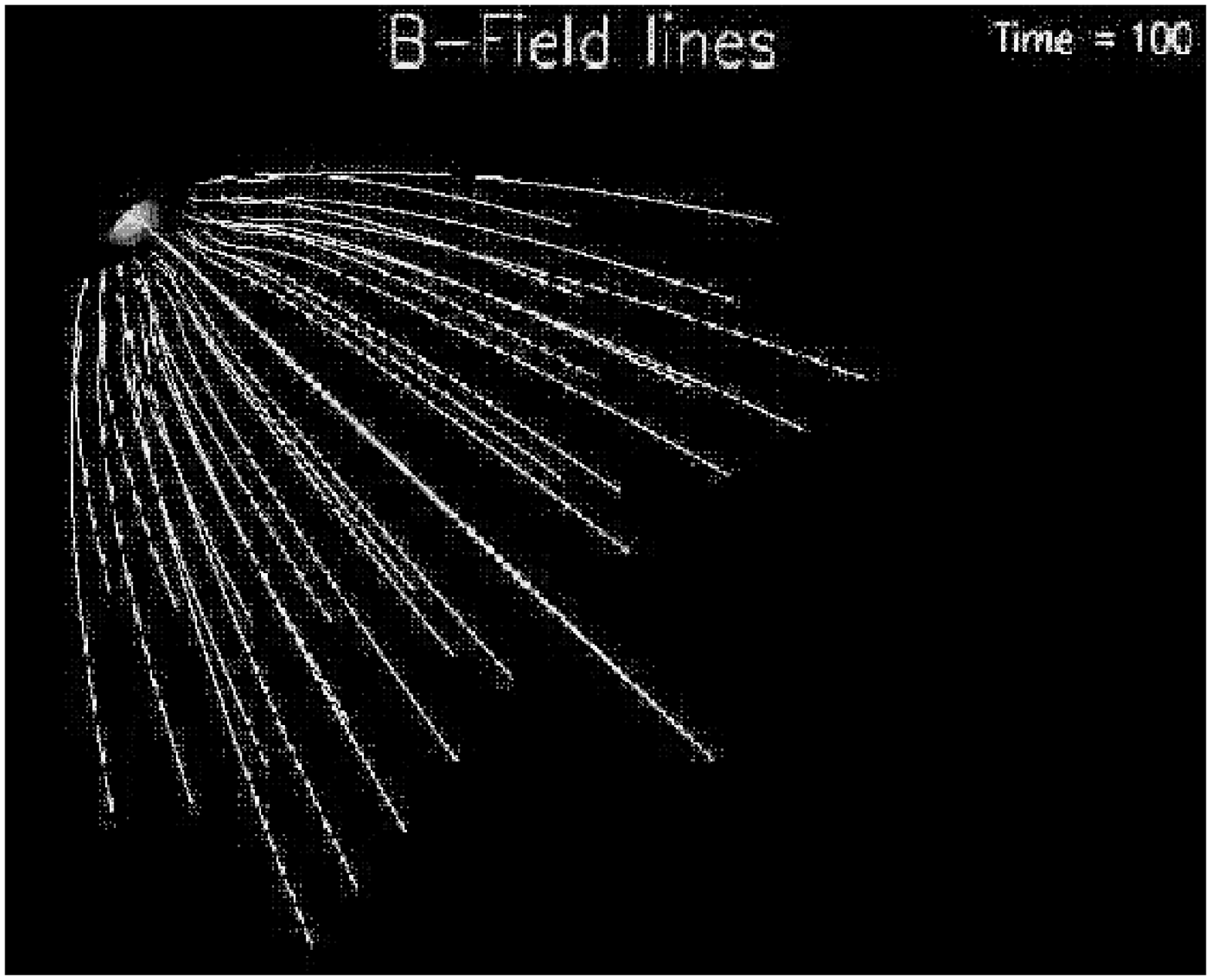}}
 \caption{{\bf Current-free solutions ($B_{\phi}=0$)} - the open
case ($\mu = -1$): Shown here are the magnetic
field lines at $\tau=0$ (top) and 
at $\tau=100$ (bottom).
In this figure and the rest of figures, 
the blob in the center represents an iso-surface density
around the central object (the proto-star is less
than few pixels in size and cannot be seen in the figure).
The  rotation axis is shown crossing the iso-surface density into the central
object and is in the plane of the paper. The accretion
disk not shown here is to the far left
and is perpendicular to the rotation axis.}
\end{center}
 \label{Figure 1.}
\end{figure}

\begin{figure}[t!]
\begin{center}
 \resizebox{6cm}{!}{\includegraphics{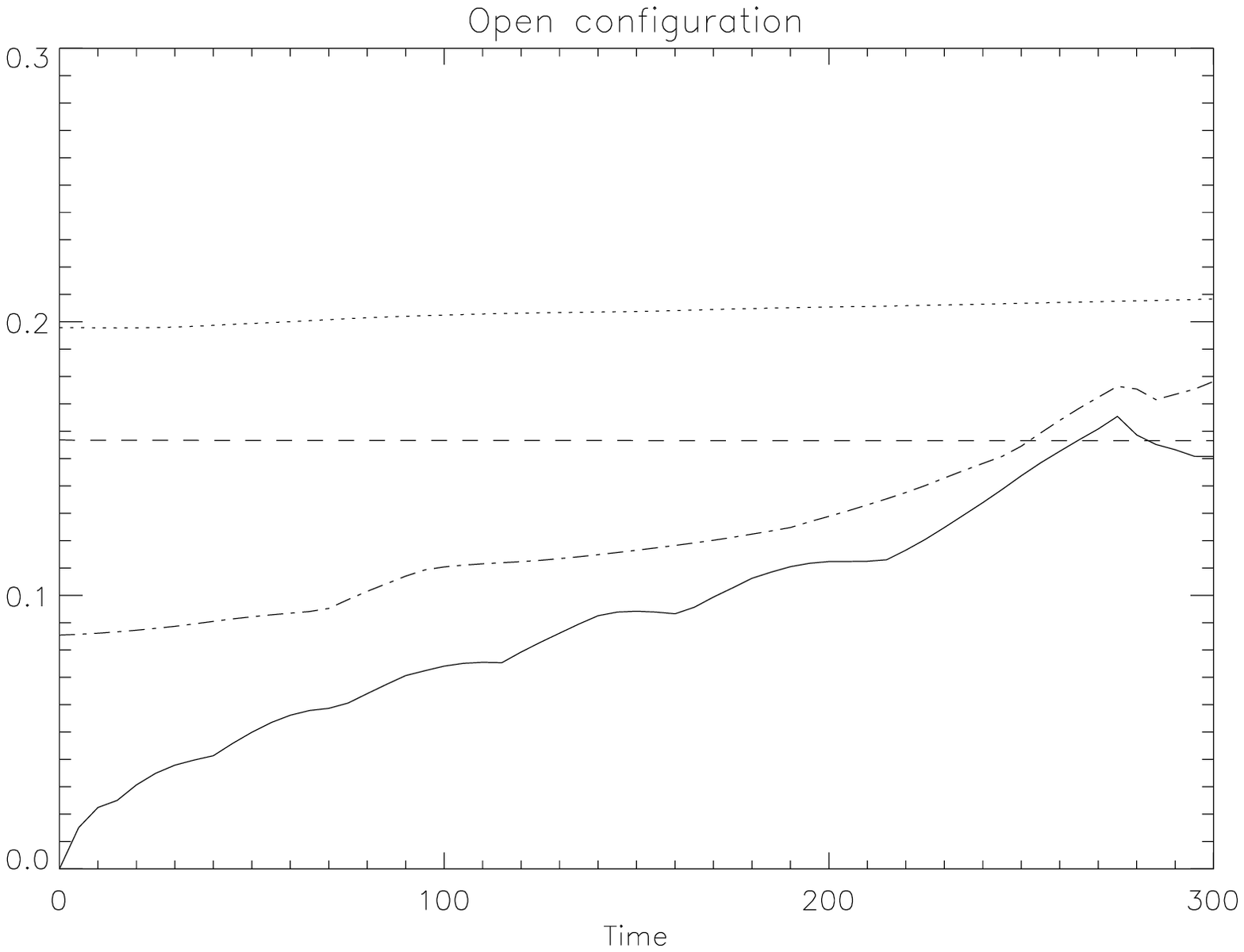}}
 \resizebox{6cm}{!}{\includegraphics{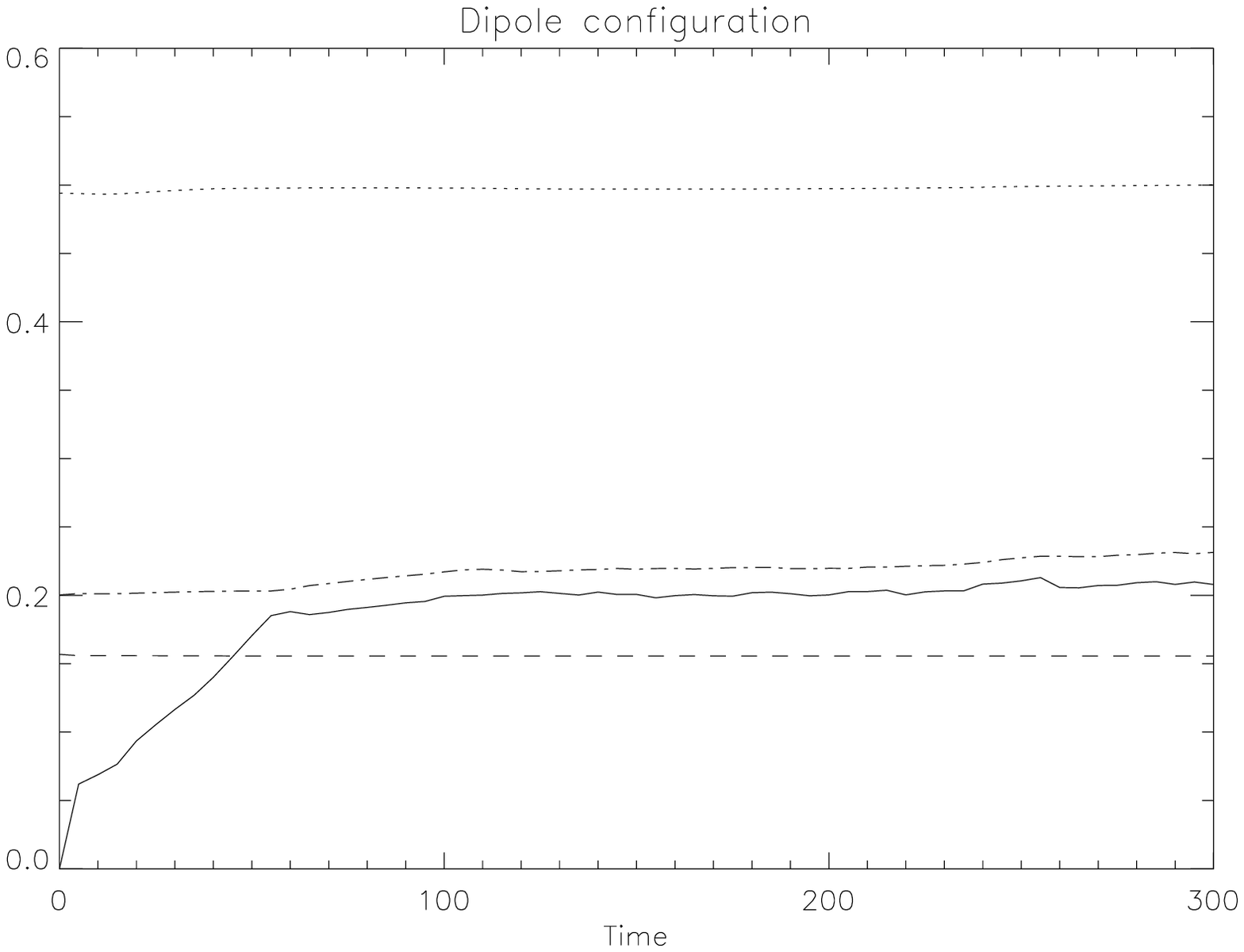}}
 \caption{{\bf Current-free solutions ($B_{\phi}=0$)} - the open
case ($\mu= -1$; top) and the dipole
case ($\mu = -2$; bottom). Shown evolving in time 
are the maximum values for the velocity (solid line), the magnetic field
strength (dotted line), and the density (dashed lines). The dot-dashed line
shows the maximum speed for the propagation of Alfv\'en waves.}
 \label{Figure 2.}
\end{center}
\end{figure}

Cartesian coordinates, $(x,y,z)$, are used for all simulations. 
While being the natural system to use to avoid any directional biases, it does
introduce some of its own problems not encountered in the 2-D cylindrically
symmetric simulations. We refer
the interested reader to OPC for the technical
reasons underlying the choice of Cartesian coordinates for such simulations.
The disc is taken to lie along the $x$--$y$ plane, and the disc axis
corresponds to the $z$-axis.  In units of the inside radius of the disc,
$r_i$, the simulated region has dimensions $(-15:+15, -15:+15,
0:+60)$, and is divided into $(95, 95, 120)$ uniform rectangular zones.

\begin{figure}[t!]
\begin{center}
 \resizebox{6cm}{!}{\includegraphics{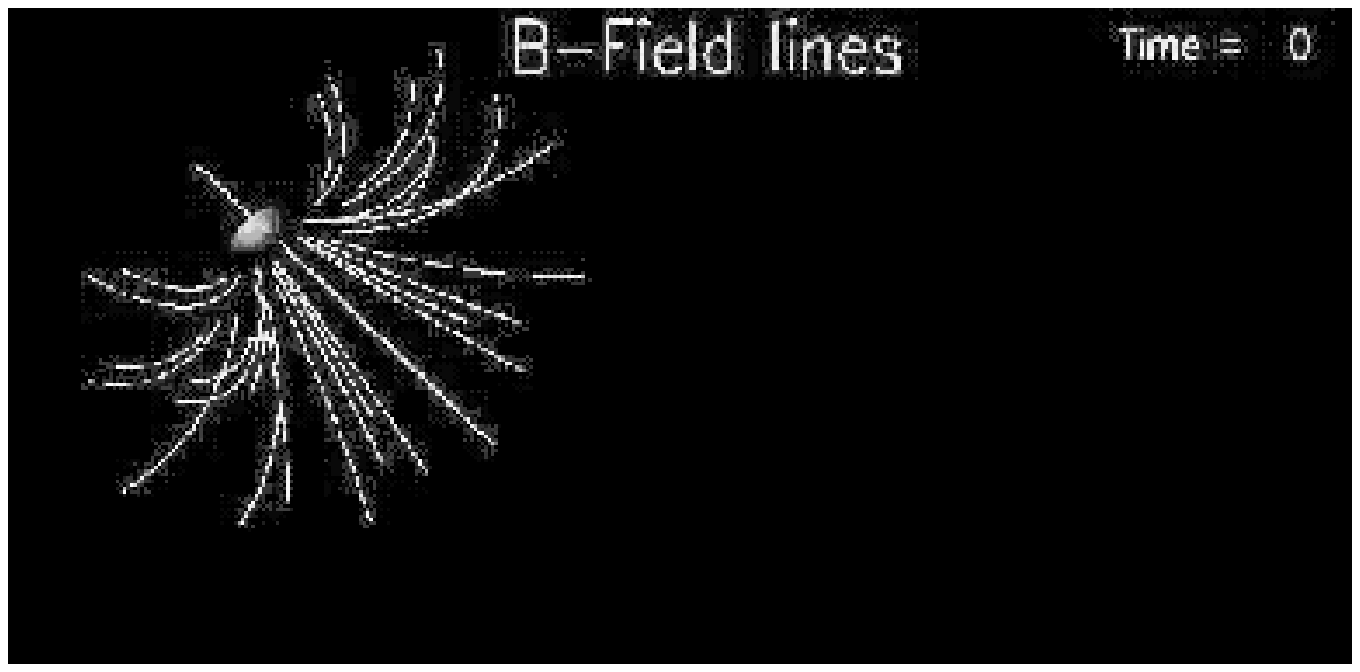}}
 \resizebox{6cm}{!}{\includegraphics{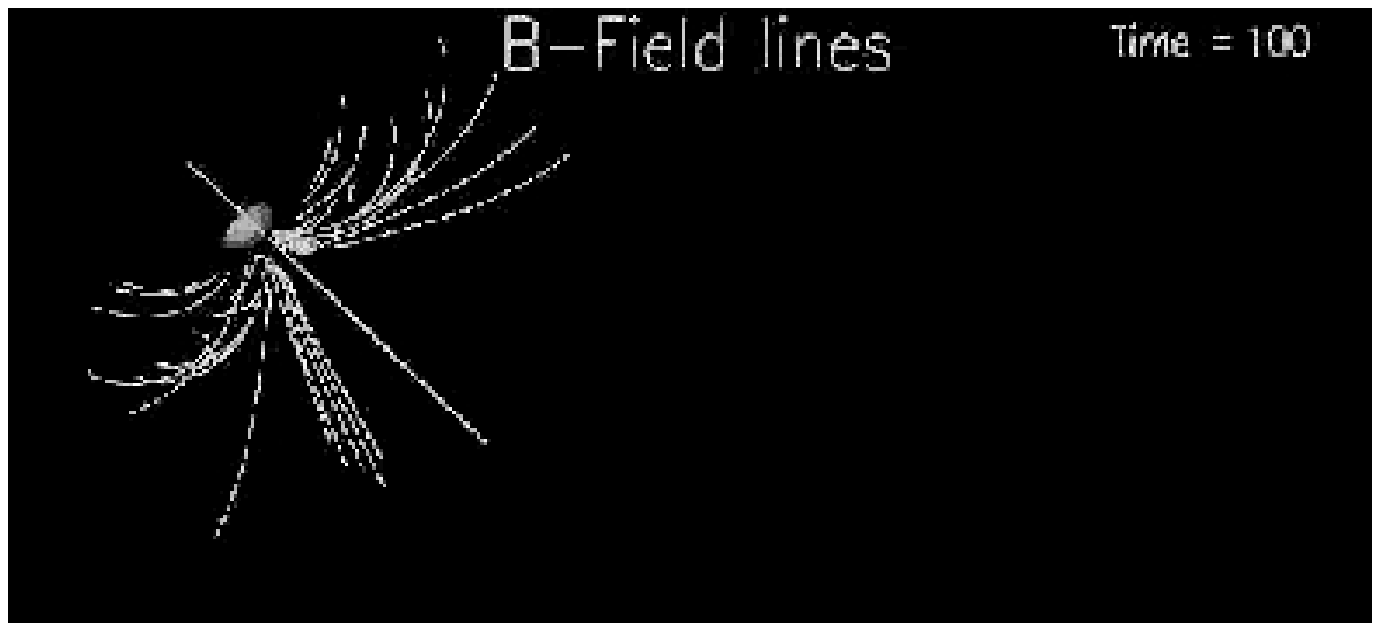}}
 \caption{{\bf Current-free solutions ($B_{\phi}=0$)} - the dipole
case ($\mu = -2$). Shown  are  the magnetic
field lines at $\tau=0$ (top) and 
at $\tau=100$ (bottom).}
\end{center}
 \label{Figure 3.}
\end{figure}

Fig.1 shows the magnetic field lines for the open
configuration at time $\tau =0$ (in units
of inner Kepler period) and at time $\tau = 100$.
The magnetic configuration changes only slightly in time and the generated 
currents are negligible. 
This is further demonstrated in Fig.2 (top panel) where
we show the time evolution of the maximum values
for velocity, density and magnetic field strength. 
The induced velocity and slight variations in the
magnetic field topology are signature of Alfv\'en waves which
 as expected have no dynamical effects on the coronal material.
The solution stays force-free up to
$\tau =300$ and no motion/jet is generated - as can 
further be checked from the movies of the simulations (at
{\it http://www.fys.ku.dk/$^{\sim}$svart/Jets/}). 
The closed configuration (lower panel in Fig.2) is also
stable in time  although 
the induced  errors are 20\% larger than in the open case (Fig.3). 
Here as well the initial hydrostatic balance
is preserved up to $\tau= 300$.

The simulations described above are easily implemented 
numerically and if it is left unperturbed, the corona will remain in perfect
numerical  balance  to within machine round-off errors.
In any case, when disk rotation and mass-load are taken into account,
the small perturbations are quickly and completely washed out by the jet
dynamics  ($\tau \ll 50$). The complete set of simulations
with and without disk rotation and mass-loading can be
visualized and compared at {\it http://www.fys.ku.dk/$^{\sim}$svart/Jets/}.

\subsection{The {\bf JETSET} tool}

We developed a tool, namely {\it JETSET}, that 
generates initial
states as described above. 
The {\it JETSET} main frame is shown in Fig. 4. 
Once the grid dimensions, the physical scales and the appropriate
coronal and disk parameters have been specified, 
{\it JETSET} performs a Newton-Raphson method to find the
correct density distribution while the corresponding magnetic
field configuration is computed using the approach described
in previous sections. 
The resulting Data (density, specific energies, 
velocity, and magnetic field) describing the initial setup
is stored in an HDF (Hierarchical Data Format) file  which 
can then be read by the user's code (such as ZEUS).
The magnetic field lines can be visualized (see Fig. 4)
as well as the density distribution around the central object (see Fig. 5).
{\it JETSET} 
is available (down-loadable) at 
{\it http://www.nordita.dk/$^{\sim}$ouyed/JETTOOLS/}.
Included in the package are {\it README} and {\it HELP} files.

\section{Conclusion}

In this paper, we solved analytically for force-free 
solutions of magnetic configurations
which can be implemented in 3-D simulations
of astrophysical jets (disk winds). These configurations
which thread the accretion disk and the corona, we showed, do not 
perturb the initial hydrostatic balance
and are stable in time. 
  While idealistic
(developed for ease of implementation and computation),
they constitute the first stage towards testing the effects of
different magnetic configurations on the simulated jets
in 3-D.  Realistic configurations ought to reproduce basic
features of astrophysical jets, such as their cylindrical shape,
their knotty structure, and their stability.

\begin{acknowledgements}
We thank R. E. Pudritz and C. Rogers for helpful discussions.
\end{acknowledgements}

\begin{figure*}[p]
\begin{center}
\resizebox{15cm}{!}{\includegraphics[angle=270]{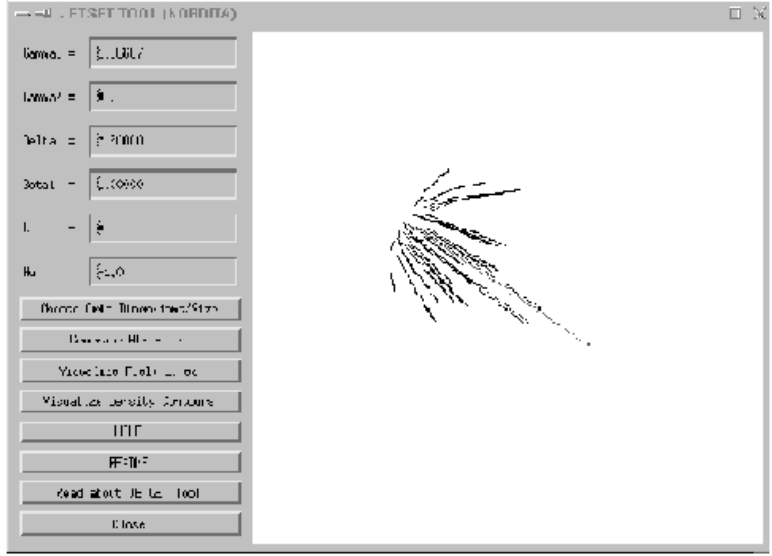}}
\caption{{\bf JETSET TOOL}: {\it JETSET} sets initial states for astrophysical
jet simulations in 3-D.  
The initial set up consists of a corona in hydrostatic
 balance around a central object with current-free magnetic fields threading
through the corona and the underlying accretion disk (fixed boundary).
{\it JETSET} solves for the correct density distribution around
the central object and the appropriate current-free magnetic field configuration
once the two parameters, $k$ and $\mu$, are specified (see text).
The coronal material  might consist of one pressure component ($\gamma_1 = 5/3$,
$\gamma_2 =0$) or two-pressure component ($\gamma_1\ne 0$, $\gamma_2\ne 0$,
where $\gamma_1$ and $\gamma_2$ are the corresponding adiabatic indices).
The initial magnetic field configuration can be visualized and saved
into a file. Figures similar to the upper panels in Fig. 1 and Fig. 2 
can be generated (and
saved into files) by {\it JETSET} as evident from the renderer to the right.
Further details on {\it JETSET} can be found in the
README and HELP files included in the {\it JETSET} package which
can be down-loaded at {\it http://www.nordita.dk/$^{\sim}$ouyed/JETTOOLS/}.}
\label{Figure 4a.}
\end{center}
\end{figure*}

\begin{figure*}[p]
\begin{center}
 \resizebox{6cm}{!}{\includegraphics[angle=270]{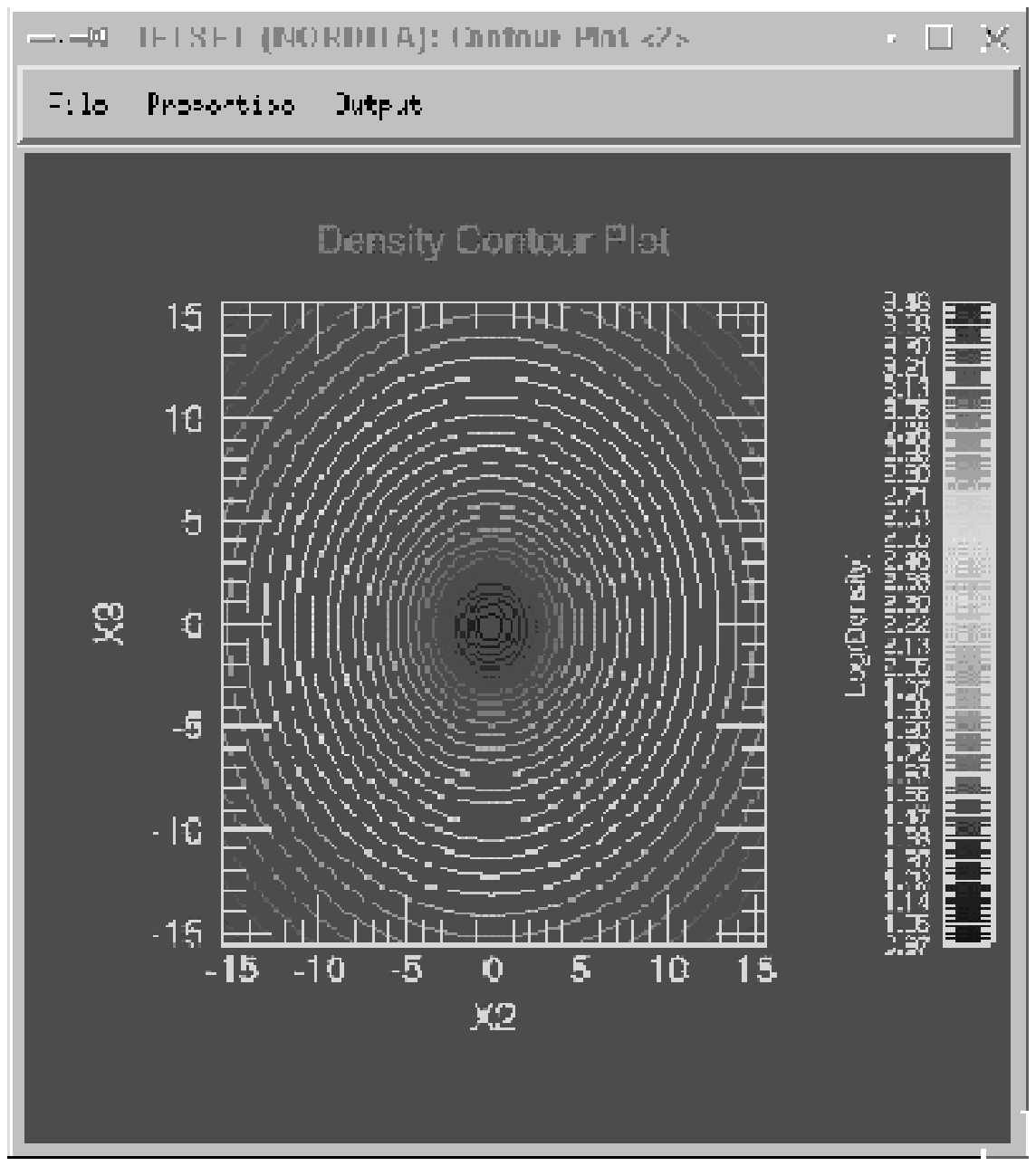}}
 \resizebox{6cm}{!}{\includegraphics[angle=270]{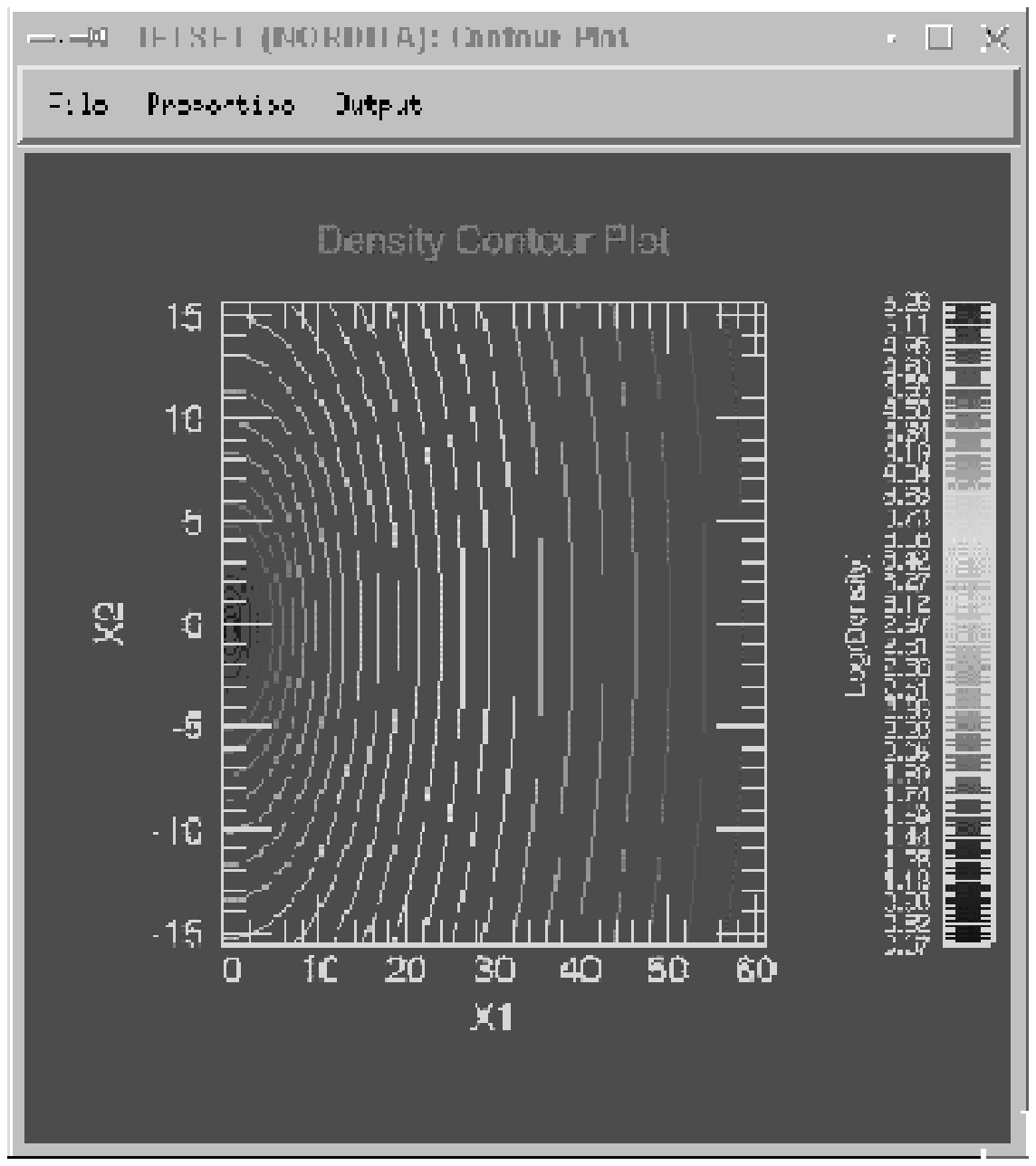}}
 \caption{{\bf JETSET TOOL} - Density contours
as generated by {\it JETSET} for a single pressure
component corona with $\gamma_{1}=5/3$. The left panel is  in a plane 
parallel to the disk surface while the  panel to the right
shows the density contours in a plane perpendicular to the
disk surface and containing the  
disk rotational axis.}
 \label{Figure 4b.}
\end{center}
\end{figure*}


\appendix
\section{Current-free  configurations}\label{cfapp}

\subsection{$B_{\phi}\ne 0$}

Here, the problem reduces to finding a
 general solution to Laplace's equation:
\begin{equation}
\vec{\nabla}^2\varphi = 0.
\end{equation}
Making a separation of
variables in cylindrical coordinates,
\begin{equation}
\varphi = R_\varphi(r)\Phi_\varphi(\phi)Z_\varphi(z)\ ,
\end{equation}
we find (Gradshteyn \& Ryznik),
\begin{align}
\varphi (r,\phi,&z) = \int_0^\infty\sum_{\rho=0}^\infty D(\rho,\lambda)
  J_\rho(\lambda   r)  A(\rho,\lambda) e^{-\lambda |z|}
 \times \nonumber \\
& \left ( E(\rho,\lambda)  \sin(\rho \phi) +
C(\rho,\lambda)\cos(\rho\phi) \right )\textrm{d}\lambda
 \label{scalarpot2},
\end{align} 
where $\lambda$ and $\rho$
are constants of separation, and $J_\rho(\lambda   r)$ the
Bessel function. The unknown
functions $A, C, D, E$ which can in fact be combined 
in two constants (see below) within the Fourier space are to be determined from the
boundary conditions. Equation \eqref{scalarpot2} is a superposition of the
solutions to $R_\varphi, \Phi_\varphi$ and $Z_\varphi$. It shows that the
constants may be determined from a single boundary condition, for example the
configuration of the toroidal magnetic field in the disk $B_{\phi,0} =
B_\phi(r,\phi,0)$.

We now look at the $B_\phi = \vec{\nabla}\varphi|_\phi$ component:
\begin{eqnarray}\label{solphi}
B_\phi = \sum_{\rho=0}^\infty &&\left[ \cos(\rho \phi)\left ( \rho
\int_0^\infty S_{1,\rho}(\lambda) 
e^{-\lambda |z|} J_\rho(\lambda   r) \textrm{d}\lambda \right ) + \right.
\nonumber \\
&& \left. \sin(\rho\phi) \left. \left (- \rho \int_0^\infty 
S_{2,\rho}(\lambda) e^{-\lambda |z|}  J_\rho(\lambda   r) \textrm{d}\lambda 
\right ) \right.\nonumber \right ]\ .
\end{eqnarray}
where  $S_{1,\rho}(\lambda)= D(\rho,\lambda)  E(\rho,\lambda)  A(\rho,\lambda)$ and
$S_{2,\rho}(\lambda)= D(\rho,\lambda) C(\rho,\lambda)   A(\rho,\lambda)$

Equation above is simply a Fourier expansion in the toroidal
 dependence of the magnetic field. The integrals before the trigonometric
functions are the Fourier coefficients; so we may write :
\begin{eqnarray}
\int_0^\infty S_{1,\rho}(\lambda) \lambda   e^{-\lambda |z|}
 J_\rho(\lambda   r) \textrm{d}\lambda &=&
\frac{1}{\pi\rho}\int^{2\pi}_{0}B_{\phi}\sin(\rho \phi)
\textrm{d}\phi,\nonumber \\ 
-\int_0^\infty S_{2,\rho}(\lambda) \lambda   e^{-\lambda |z|}  J_\rho(\lambda
  r) \textrm{d}\lambda &=& \frac{1}{\pi\rho}\int^{2\pi}_{0}B_{\phi}\cos(\rho
\phi) \textrm{d}\phi \nonumber,
\end{eqnarray}
 where the right hand side is  the
expressions for the Fourier coefficients. 
Note that $S_{1,\rho}(\lambda)$ and $S_{2,\rho}(\lambda)$ are the Hankel
transforms of the Fourier components of $B_{\phi}(r,\phi,0)$. 
That is, for any given toroidal configuration of the magnetic field in the
disk, $S_{1,\rho}(\lambda)$ and $S_{2,\rho}(\lambda)$ are given by equations
above  thus providing a simple procedure for determining the scalar field.

Finally, note that the simplest magnetic
 configuration possible is given by $S_\rho(\lambda)=0 \, \forall \, \rho
\neq k$ which gives us the boundary condition investigated in this paper
$B_{\phi,0} = (b\ r^{\mu-1})\times \sin (k\phi)$ (including the $k=0$ case with
no toroidal dependence); $b$ is  a scaling factor.

\subsection{$B_{\phi} = 0$}

The corresponding scalar field  is obtained by cancelling the
$\phi$ dependence in \eqref{scalarpot2}. The components $B_{r}$
and $B_{z}$ are then direct derivative of $\varphi$. For example,
one can show that
\begin{equation}
B_{z} = \int_0^\infty S(\lambda) \lambda J_0 (\lambda r) e^{-\lambda |z|}
\textrm{d}\lambda \nonumber\ ,
\end{equation}
where $S(\lambda)$ is the Hankel transform of $B_{z}(r,\phi,0)$.

\end{document}